\documentclass{moriond}

\usepackage{amsmath,amssymb}
\usepackage{graphics,epsfig,wrapfig,times}

\def\be{\begin{equation}}
\def\ee{\end{equation}}
\def\bea{\begin{eqnarray}}
\def\eea{\end{eqnarray}}

\newcommand{\Photo}{\includegraphics[height=35mm]{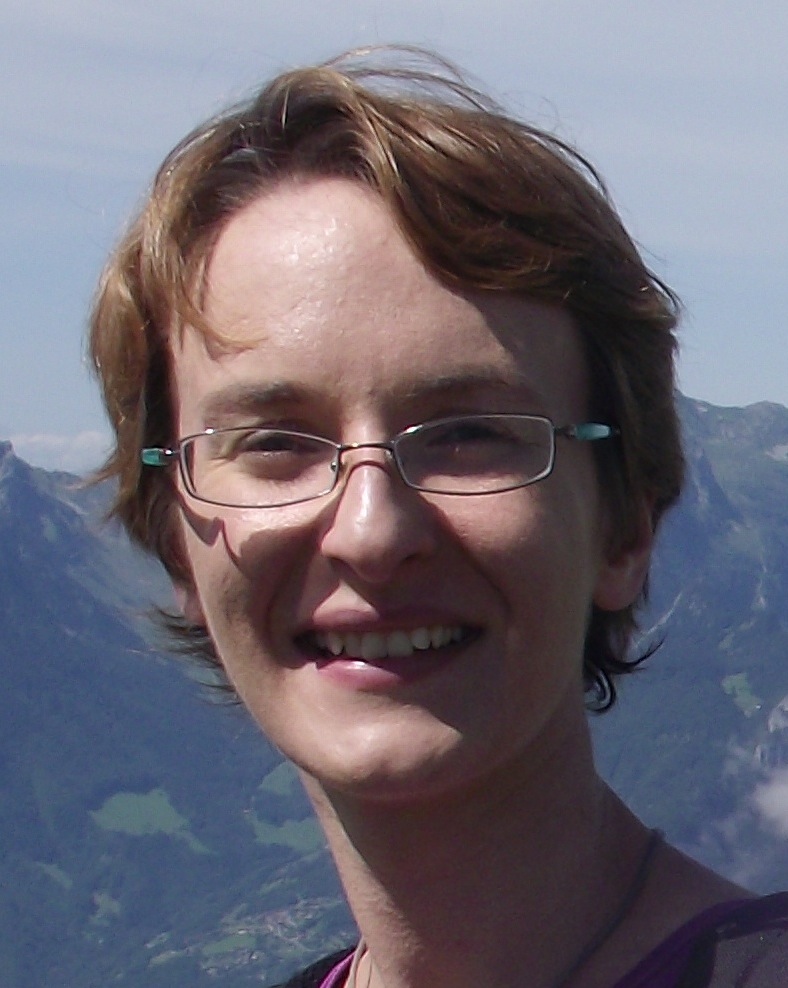}}

\begin{document}

\vspace*{-2.0cm}
\begin{flushright}
MITP/15-035
\end{flushright}

\vspace*{4cm}
\title{Improving LHC searches for strong EW symmetry breaking resonances}

\author{Anna Kami\'{n}ska}

\address{PRISMA Cluster of Excellence and Mainz Institute for Theoretical Physics\\
 Johannes Gutenberg University, 55099 Mainz, Germany}

\maketitle
\abstracts{
Composite Higgs models generically predict the existence of heavy spin-1 resonances with the same quantum numbers as electroweak gauge bosons. The effective lagrangian description of these resonances is presented, pointing out the origin of their interactions with SM matter fields and the resulting LHC phenomenology. Search strategies for spin-1 resonances are discussed, mentioning possible advantages offered by boosted decay products. The impact of interactions between spin-1 resonances and fermion resonances, crucial for the proper interpretation of LHC searches, is pointed out.
}

\section{Introduction}

Composite Higgs models invoke electroweak symmetry breaking by new strong dynamics, in analogy with chiral symmetry breaking in QCD, without introducing the hierarchy problem. In order to be consistent with the measured properties of the Higgs boson and the non-observation of new particles beyond the Standard Model, the composite Higgs is a pseudo-Nambu-Goldstone (PNG) boson of some global symmetry of the new strong sector, similar to the pion of QCD. The minimal composite Higgs model (MCHM) \cite{Agashe:2004rs} is based on $SO(5)\rightarrow SO(4)\sim SU(2)_L\times SU(2)_R$ global symmetry breaking, inducing a full pseudo-Goldstone Higgs doublet.

In the effective Lagrangian description, which will be used throughout this note, the PNG bosons $\Pi\left( x\right) =\Pi^{\hat{a}}\left( x\right) T^{\hat{a}}$ of $SO(5)\rightarrow SO(4)$ are parameterized by $U\left( \Pi\right) =e^{i\Pi\left( x\right) /f}$ transforming as
\begin{equation}
U\left( \Pi\right) \rightarrow g\; U\left( \Pi\right) \; h^{\dag}\left( \Pi , g\right) ,\ \ \ \ \ \ g\in SO(5),\; h\in SO(4).
\end{equation}
The leading order effective Lagrangian term describing self-interactions of these bosons takes the form
\begin{equation}
\mathcal{L}^{\Pi }=\frac{f^2}{4} Tr \left\lbrace d_{\mu}d^{\mu}\right\rbrace 
\end{equation}
where $d_{\mu}$ is defined by
\begin{equation}
-iU^{\dag}D_{\mu}U=d^{\hat{a}}_{\mu}T^{\hat{a}}+E^{a}_{\mu}T^{a}=d^a+E^a 
\end{equation}
and $T^{\hat{a}}$, $T^a$ are respectively the broken and unbroken generators of $SO(5)$. The covariant derivative takes into account the external gauging and introduces interactions of the composite Higgs doublet with elektroweak bosons. Whenever the generators corresponding to the electroweak group are "missaligned" with the generators of $SO(4)$, the electroweak symmetry is broken. Such "missalignement" is induced by loop corrections related to $SO(5)$ violating Yukawa and gauge interactions. Naturalness of electroweak symmetry breaking requires that the fermion resonances accompanying SM fermions in the composite Higgs framework have masses not far from the electroweak scale, $m_F\lesssim{} 1$TeV \cite{Pomarol:2012qf}.

The generically expected experimental signatures of strong electroweak symmetry breaking range from flavor physics, electroweak precision data (S and T parameters), modification of Higgs couplings to the direct observation of new states - fermion resonances and spin-1 resonances. In this note I address the phenomenology of spin-1 resonances in composite Higgs models.

\section{Spin-1 resonances in composite Higgs models}

In order to speak about direct production and LHC phenomenology of spin-1 resonences, one has to introduce them explicitly into the effective description. There are several formalisms that allow to do that, the CCWZ formalism \cite{Coleman:1969sm,Callan:1969sn} and the "hidden local symmetry" \cite{Bando:1987br} formalism being the most popular ones. At leading order in the effective description, both these formalism lead to equivalent results. The spin-1 resonances related to $\mathcal{G}\rightarrow\mathcal{H}$ symmetry breaking are expected to appear in a representation of the unbroken global symmetry of strong dynamics, in the case of MCHM - in representations of $SO(4)$. In the "hidden local symmetry" formalism the effective Lagrangian for vector mesons is constructed by enlarging the symmetry structure to $\mathcal{G}\times\mathcal{H}_{local}\rightarrow\mathcal{H}$, where $\mathcal{H}_{local}$ is the "hidden" gauge group (a purely mathematical tool, without physical meaning). The "hidden local symmetry" is broken and its heavy gauge bosons $\rho^{\mu}$ provide degrees of freedom for the effective description of spin-1 resonances. The effective description of a minimal composite Higgs model with a single set of spin-1 resonances has, at the leading-order Lagrangian level, only three free parameters, which can be chosen as
\begin{equation}
m_{\rho},\ g_{\rho},\ \xi=\frac{v_{EW}^2}{f^2}
\end{equation}
where $m_{\rho}$ is the mass of the set of vector resonances, $g_{\rho}$ is the "hidden" gauge coupling and $\xi$ describes the hierarchy between the weak scale and the energy scale of strong dynamics $f$. Parameter $\xi$ also scales the departure from SM values of Higgs couplings to electroweak gauge bosons, hence it is restricted to be small $\xi\lesssim 0.1$ by experimental data. The interactions of spin-1 resonances with the Higgs boson are set by the symmetry breaking structure and the PNG-boson nature of the Higgs. The interactions of $\rho$ resonances with SM matter fields are induced by two effects
\begin{enumerate}
\item mass mixing between $\rho^{\mu}$ and $W^{\mu},\; Z^{\mu}$ fields; this automatically introduces interactions of $\rho$ resonances with electroweak gauge bosons and interactions with SM fermions through mass mixing effects feeding into the covariant derivatives in fermion kinetic terms
\item direct interactions between $\rho^{\mu}$ and fermion resonances, which mass mix with SM fermions (partial compositeness); the description of this effect requires introducing fermion resonances into the effective Lagrangian, which is model dependent.
\end{enumerate}
Naive dimensional analysis (NDA) suggests a connection between the parameters introduced above, $m_{\rho}\sim g_{\rho} f$. In order to present a more general picture, I will treat them for the time being as idependent parameters. In the following I consider only a single set of spin-1 resonances transforming in the adjoint representation of $SU(2)_L$, as on grounds of general arguments such resonences are most likely to be most relevant for LHC searches.

\subsection{$\rho$ properties from mass mixing effects in the electroweak sector}

\begin{figure}
\begin{minipage}{0.5\linewidth}
\centerline{\includegraphics[width=0.7\linewidth]{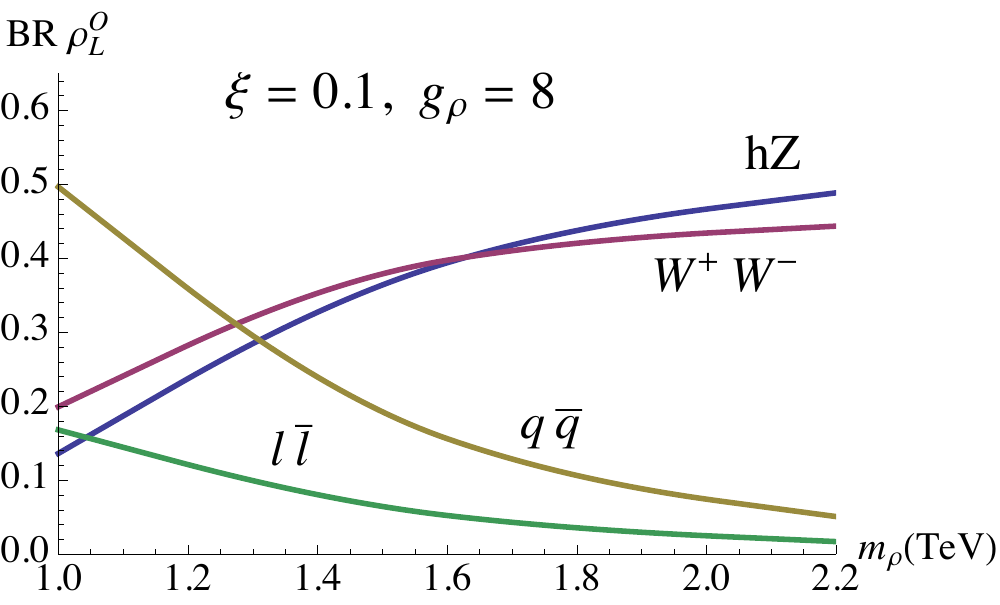}}
\end{minipage}
\hfill
\begin{minipage}{0.5\linewidth}
\centerline{\includegraphics[width=0.7\linewidth]{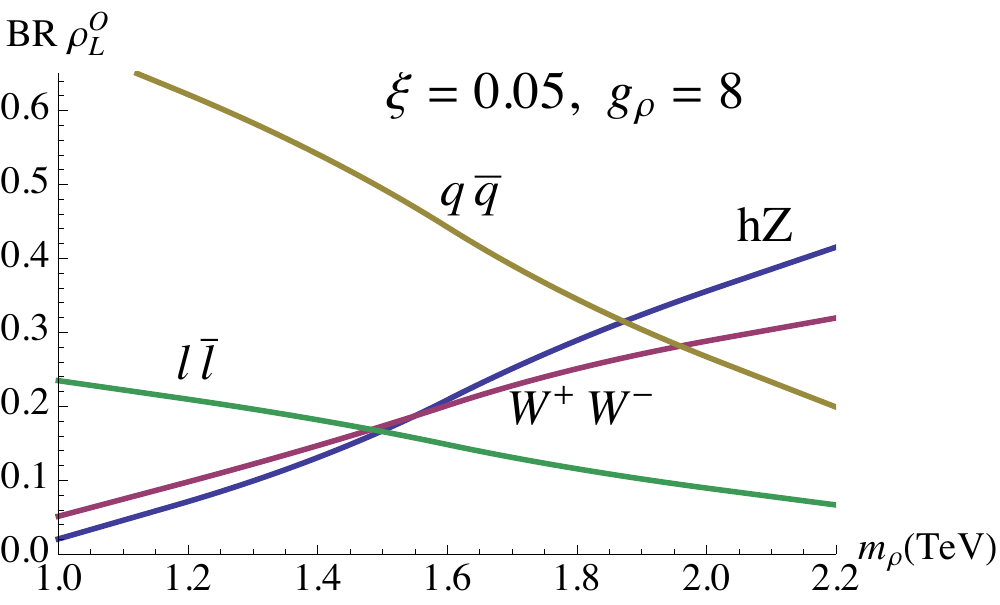}}
\end{minipage}
\caption[]{Branching ratios of $\rho$ resonances from $\rho^{\mu}$ and $W^{\mu},\; Z^{\mu}$ mass mixing effect.}
\label{SU2xi005g8BR}
\end{figure}

\begin{figure}
\begin{minipage}{0.5\linewidth}
\centerline{\includegraphics[width=0.8\linewidth]{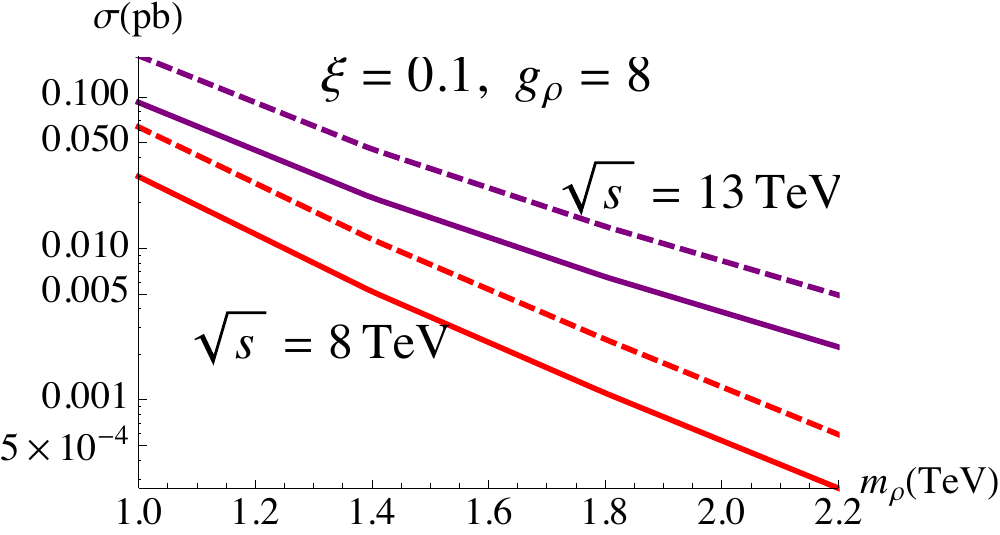}}
\end{minipage}
\hfill
\begin{minipage}{0.5\linewidth}
\centerline{\includegraphics[width=0.8\linewidth]{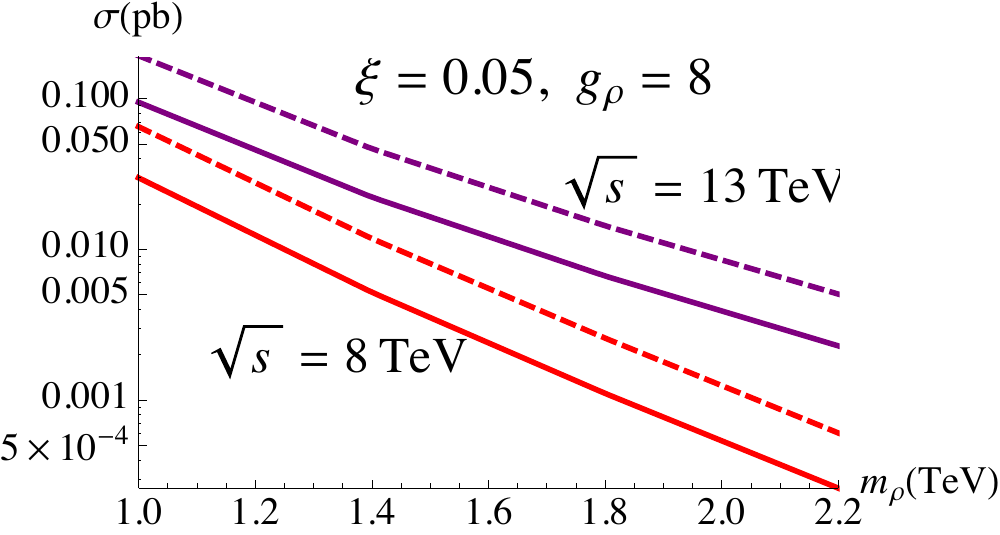}}
\end{minipage}
\caption[]{Production cross-sections of $\rho$ resonances from $\rho^{\mu}$ and $W^{\mu},\; Z^{\mu}$ mass mixing effect.}
\label{SU2xi005g8DY}
\end{figure}

Let us first consider the effect of mass mixing between $\rho^{\mu}$ and $W^{\mu},\; Z^{\mu}$ fields alone. This leads to the following decay widths for spin-1 resonances
\begin{eqnarray}
\Gamma\left( \rho^0\rightarrow W^+ W^-\right) & \approx & \Gamma\left( \rho^0\rightarrow Z h\right) \approx \frac{m_{\rho}^5 \xi^2}{192 \pi g_{\rho}^2 v^4}.
\nonumber \\
\Gamma\left( \rho^0\rightarrow e^+ e^-\right) & \approx & \Gamma\left( \rho^0\rightarrow \mu^+ \mu^-\right) \approx  \frac{g^4 m_{\rho} \left( 1+\sqrt{1-\xi}\right) ^2}{96\ 4\pi g_{\rho}^2} \nonumber \\
\Gamma\left( \rho^0\rightarrow q_i \bar{q_i}\right) & \approx & \frac{g^4 m_{\rho} \left( 1+\sqrt{1-\xi}\right) ^2}{32\ 4\pi g_{\rho}^2} .
\end{eqnarray}
The mixing angle between $\rho^{\mu}$ and $W^{\mu},\; Z^{\mu}$ fields is proportional to $g/g_{\rho}$, hence it is not surprising that the deacy widths are suppressed by $1/g_{\rho}^2$. For sufficiently large values of $m_{\rho}$ decays into gauge boson pairs and a gauge boson plus a Higgs boson will always dominate. However, for small values of $\xi$ the $\rho$ resonance decays into fermion pairs are non-negligible, especially in the low mass region. This can be seen in figure \ref{SU2xi005g8BR} for a specific value of $g_{\rho}=8$ and two values of $\xi$, $\xi=0.1$ (left) and $\xi=0.05$ (right).
The LHC production of $\rho$ resonances is dominated by Drell-Yan $q\bar{q}\rightarrow \rho$. The production cross-sections for LHC@8TeV and LHC@13TeV are presented in figure \ref{SU2xi005g8DY} as a function of $m_{\rho}$, again for a specific value of $g_{\rho}=8$ and two values of $\xi$, $\xi=0.1$ (left) and $\xi=0.05$ (right). The dashed lines correspond to the production of charged resonances, while the solid lines are represent the production of neutral resonances.

\begin{figure}
\begin{minipage}{0.5\linewidth}
\centerline{\includegraphics[width=0.8\linewidth]{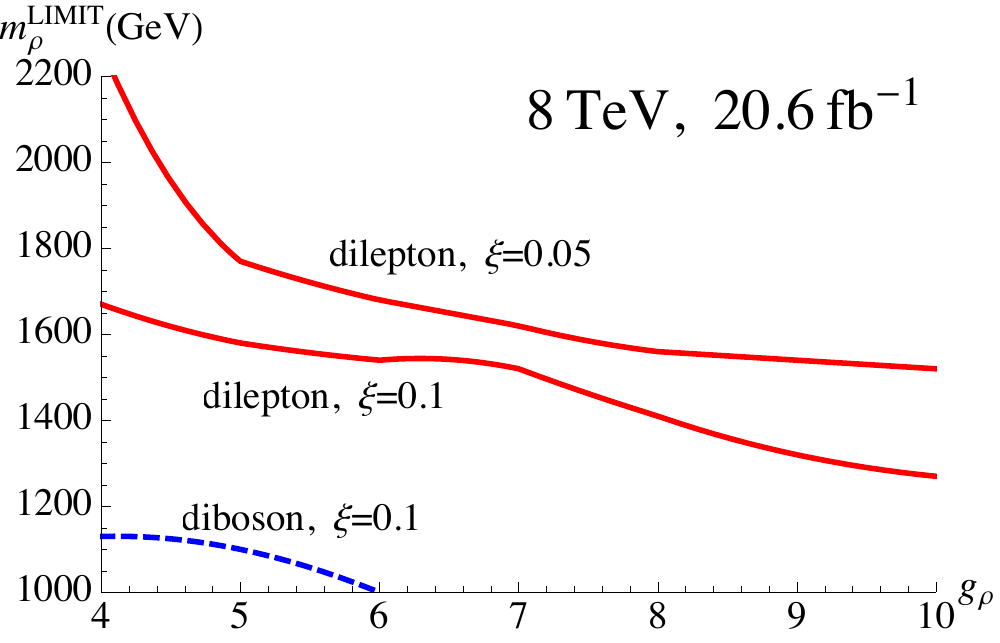}}
\end{minipage}
\hfill
\begin{minipage}{0.5\linewidth}
\centerline{\includegraphics[width=0.8\linewidth]{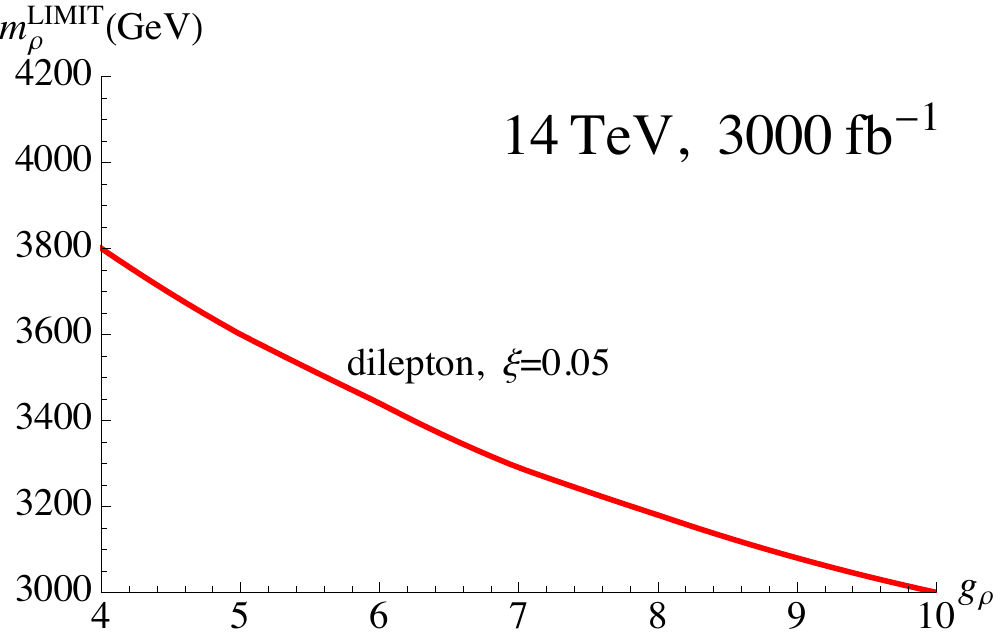}}
\end{minipage}
\caption[]{Exclusion limits for $\rho$ resonances from $\rho^{\mu}$ and $W^{\mu},\; Z^{\mu}$ mass mixing effect.}
\label{limpredSU2xi005}
\end{figure}

LHC exclusion limits for the spin-1 resonance mass can be obtained by using the publicly available search results for diboson, dilepton and dijet resonances. Presently the most stringest constraints are given by CMS dilepton resonance searches \cite{dileptonCMS} and are presented in figure \ref{limpredSU2xi005} (left) in the $g_{\rho}-m_{\rho}$ plane for specific values of $\xi=0.1,\; 0.05$. The right panel of figure \ref{limpredSU2xi005} shows the predicted range of these exclusion limits for LHC@14TeV.
The present day exclusion limits on the mass of spin-1 resonances are slightly below 2TeV, while the predicted future LHC reach goes somewhat above 3TeV.

In order to maximize the LHC potential for the discovery of spin-1 resonances one should also make use of the decay channel $\rho\rightarrow Vh$, which becomes dominant in the high $\rho$ mass region together with decays to gauge boson pairs. Together with M.Hoffmann R.Nikolaidou and S.Paganis we have looked at the potential impact of an LHC search for heavy vector mesons decaying to an electroweak gauge boson and a Higgs boson, using the fact that the Higgs boson will be highly boosted \cite{Hoffmann:2014aha}. Using a $ p_{\perp}\geq$ 550 GeV cut on the transverse momentum of the Higgs system in very clean $h \rightarrow \gamma \gamma $ and $h \rightarrow ZZ^{(*)} \rightarrow 4 \ell$ (where $\ell=e,\mu$, $V \rightarrow jj$) decay channels allows for significant reduction of the SM background. This method can lead to exclusion limits on $\rho$ mass $\sim 3$TeV at LHC@14TeV.

\subsection{The impact of interactions between $\rho$ resonances and fermion resonances}

As mentioned before, the properties of spin-1 resonances rely not only on the effect of $\rho^{\mu}$ and $W^{\mu},\; Z^{\mu}$ mass mixing, but also on the direct coupling between $\rho^{\mu}$ and fermion resonances,
\begin{equation}
-i\bar{\psi}g_{\rho}\gamma^{\mu}T^{a}\rho_{\mu}^{a}\psi
\label{intFR}
\end{equation}
where $T^{a}$ are generators of $\mathcal{H}$.  The coupling constant is taken to be the same as for the $\rho$ self-interaction which happens to be the case in most effective models.

In the following I will present the approximate prediction of the impact of fermion resonances on the properties of $\rho$ resonances transforming in the adjoint representation of $SU(2)_L$ using a simple toy model. The $\rho$ resonance discussed in the previous section couples to the component of $\psi$ charged under $SU(2)_{L}$, which I denote by $\psi_{L}$. In general $\psi$ mixes through the mass matrix with the SM fermion fields, forming new mass eigenstates. Before electroweak symmetry breaking the SM left-handed quark doublet $q_{L}$ mixes only with $\psi_{L}$. The quark mass matrix is diagonal for the following combinations
\begin{eqnarray}
q'_{L} & = & \cos\theta_{L} q_{L}-\sin\theta_{L}\psi_{L} \nonumber \\
\psi'_{L} & = & \sin\theta_{L} q_{L}+\cos\theta_{L}\psi_{L} 
\end{eqnarray}
where the mixing angle is determined by the Yukawa structure of the specific fermion partner construction.
The quark doublet $q'_{L}$ corresponds to massless fermion eigenstates and plays the role of the SM-like left-handed quark. It has an admixture from the composite fermion sector and following eq. \ref{intFR} interacts directly with the spin-1 resonance $\rho_{L}$ through
\begin{equation}
i\bar{q}'_{L}\gamma^{\mu}\left( -ig_{\rho}\sin^2\theta_{L} T_{L}^{a}\rho_{L\; \mu}^{a}\right) q'_{L} .
\label{quarkmix}
\end{equation}
Non-zero Higgs VEV generates further mixing effects in the fermion mass matrix. From the point of view of SM-like quark couplings to spin-1 resonances however the leading order effect is given by $\sin\theta_{L}$ discussed above. In our LHC phenomenology analysis we take into account only this leading order effect.

\begin{figure}
\begin{minipage}{0.5\linewidth}
\centerline{\includegraphics[width=0.8\linewidth]{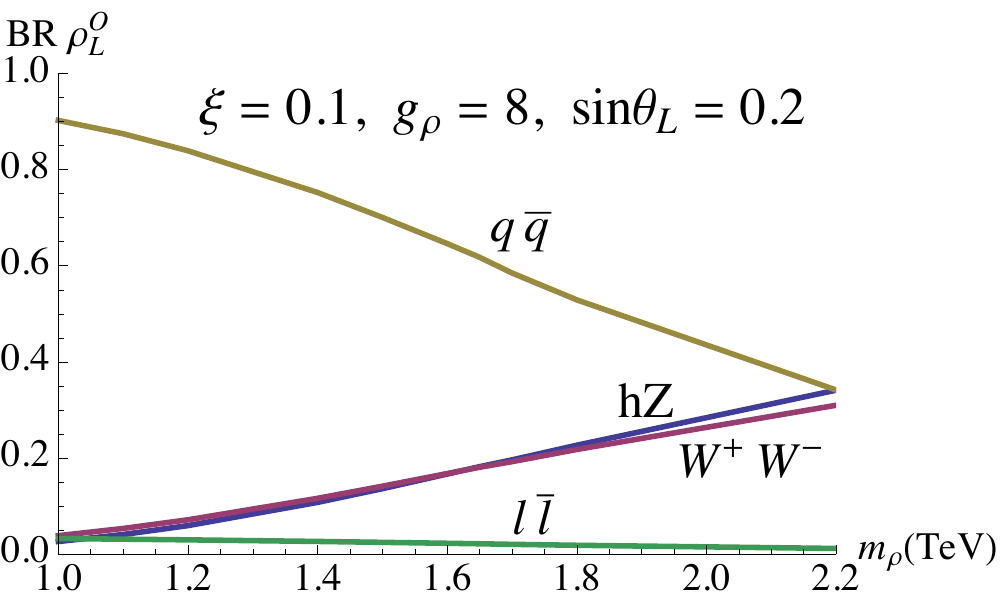}}
\end{minipage}
\hfill
\begin{minipage}{0.5\linewidth}
\centerline{\includegraphics[width=0.8\linewidth]{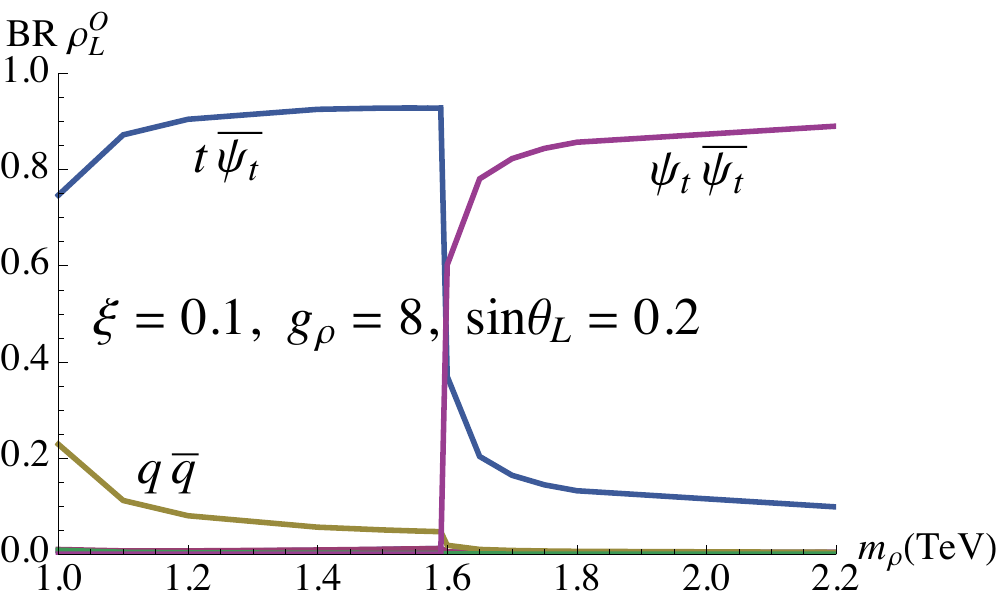}}
\end{minipage}
\caption[]{Branching ratios of $\rho$ resonances including the effect of direct coupling to fermion resonances, for $m_T\gtrsim 2$TeV (left) and $m_T\sim 0.8$TeV (right).}
\label{s2a2SU2xi01}
\end{figure}

Fermion resonances exhibit mass mixing terms with SM fermions (partial compositeness), which leads to the modification of spin-1 resonance couplings to SM fermions. This effect is especially important in the top sector, where the degree of partial compositeness is expected to be largest. If the degree of partial compositeness in first two generations of quarks is small, then the effect of direct couplings between $\rho$ resonances and fermion resonances does not affect production cross sections of $\rho$ resonances. However, the coupling of $\rho$ resonances to top quarks becomes modified, leading to the modification of branching ratios and the overall decay width. More importantly, due to the direct coupling between $\rho$ and fermion resonances, spin-1 resonances can decay into a pair fermion resonances or a fermion resonance plus a standard model fermion. The mixing in the fermion sector discussed above introduces couplings of a $\rho_L$ resonance to a light and a heavy fermion eigenstate
\begin{equation}
i\bar{q}'_{L}\gamma^{\mu}\left( -ig_{\rho}\sin\theta_{L}\cos\theta_{L} T_{L}^{a}\rho_{L\; \mu}^{a}\right) \psi '_{L} +i\bar{\psi}'_{L}\gamma^{\mu}\left( -ig_{\rho}\sin\theta_{L}\cos\theta_{L} T_{L}^{a}\rho_{L\; \mu}^{a}\right) q'_{L} .
\label{quarkmix1}
\end{equation}
along with a coupling of $\rho_L$ with two heavy fermion eigenstates scaled by $\cos^2 \theta_{L}$.
Due to naturalness arguments, top partners are expected to be light, hence the decay of $\rho$ resonances into resonances related to the top sector is highly probable. Such a new decay channel for $\rho$ can have a dramatic effect on LHC phenomenology of these resonances, on search strategies and interpretation of LHC limits. In order to show this effect let us consider two cases
\begin{enumerate}
\item heavy top partner $m_T\gtrsim 2$TeV
\item light top partner $m_T\sim 0.8$TeV.
\end{enumerate}
In the first case decays into fermion partners are not kinematically allowed in the $\rho$ resonance mass region probed presently at the LHC. 
The left panel of figure \ref{s2a2SU2xi01} shows the impact of the modification of $\rho$ couplings with third generation quarks through the coupling with fermion resonances on the branching ratios of $\rho$, for a specific choice of $\xi=0.1,\ g_{\rho}=8$ and the fermion mixing angle $\sin\theta_L=0.2$. One can notice that in the low $m_{\rho}$ region the decays of spin-1 resonances are dominated by decays into quark pairs.
Let us now see what happens when the decays into fermion partners become kinematically allowed.
The right panel of figure \ref{s2a2SU2xi01} shows the branching ratios of $\rho$, for a specific choice of $\xi=0.1,\ g_{\rho}=8$ and the fermion mixing angle $\sin\theta_L=0.2$, in the case when the top partners mass $m_T\sim 0.8$. One can see that the $\rho$ resonance decays become dominated by decays into a SM quark and its composite partner as soon as the decay becomes kinematically allowed. For larger values of $m_{\rho}$, when the decay into a pair of fermion resonances becomes possible, this decay channel quickly becomes dominant.

\begin{figure}
\begin{minipage}{0.5\linewidth}
\centerline{\includegraphics[width=0.8\linewidth]{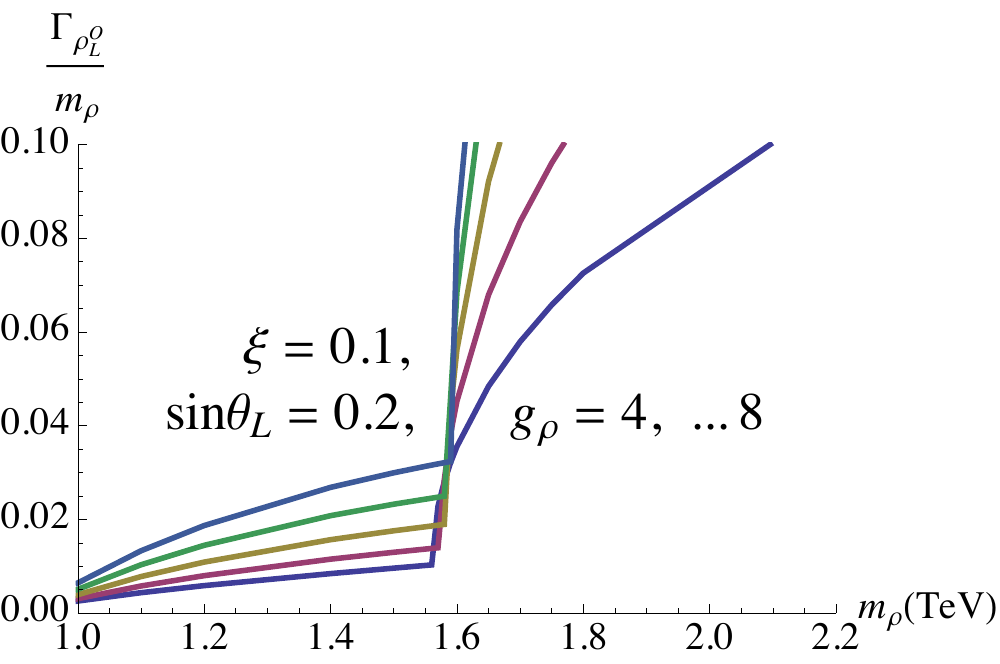}}
\end{minipage}
\hfill
\begin{minipage}{0.5\linewidth}
\centerline{\includegraphics[width=0.8\linewidth]{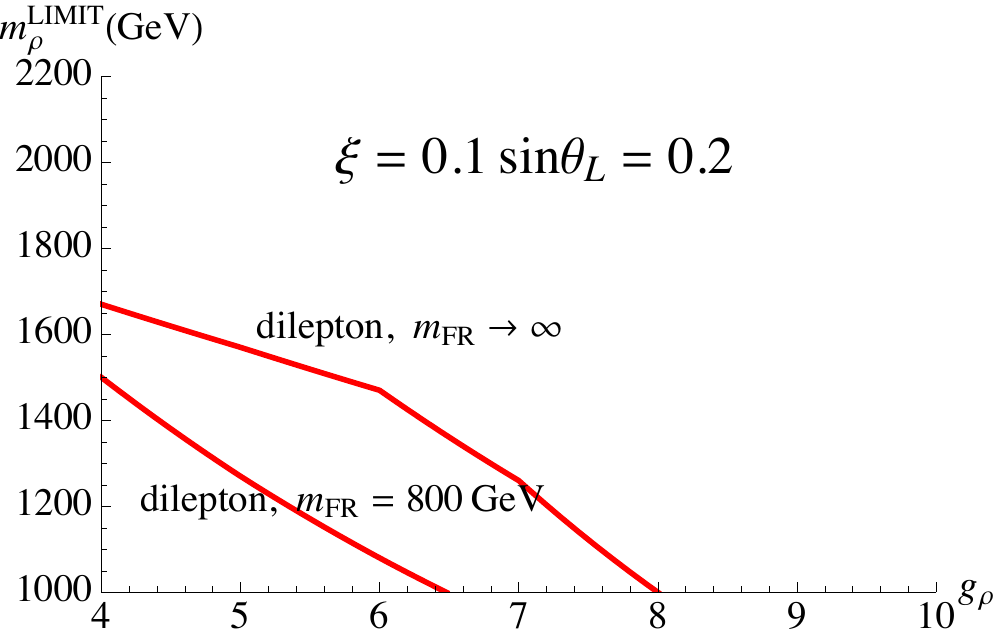}}
\end{minipage}
\caption[]{Branching ratios of $\rho$ resonances including the effect of direct coupling to fermion resonances, for $m_T\gtrsim 2$TeV (left) and $m_T\sim 0.8$TeV (right).}
\label{SU2xi01Lfr}
\end{figure}

Seeing these results, one immediately suspects that the $\rho$ resonance width is also strongly affected by decays into fermion resonances. This is in fact the case, as one can see in the left panel of figure \ref{SU2xi01Lfr} illustrating the width to mass ratio $\Gamma_{\rho}/m_{\rho}$ as a function of the resonance mass for $\xi=0.1, \sin\theta_L=0.2$ and several values of $g_{\rho}$. The overall width of $\rho$ explodes as soon as the decay into two fermion resonances becomes possible.

All these effects of the interactions between $\rho$ resonances and fermion partners modifies the intrepretation of LHC limits on $m_{\rho}$ with respect to those presented in the previous section. The limits obtained in both cases discussed above, for $m_T\gtrsim 2$TeV and $m_T\sim 0.8$TeV, are presented in the right panel of figure \ref{SU2xi01Lfr}. One can notice that the limits are weaker than previously and no limits can be set when the decays into fermion resonance pairs become kinematically available.

\section{Conclusions}

Strong electroweak symmetry breaking can be tested at the LHC in many ways - by measuring Higgs boson properties, by constraining flavor observables and by direct searches for fermion and vector resonances. In this note the effective description of spin-1 resonances and the resulting LHC phenomenology has been discussed. Decays of spin-1 resonances, for moderate values of $m_{\rho}\sim 1-2$ TeV, are often dominated by fermion pair production. Searches for dilepton resonances, due to small backgrounds, are presently sensitive to $\rho$ resonances in this mass range. LHC is already probing the parameter space of vector resonances allowed by electroweak precision data. Searches for heavy vector mesons can be improved by taking advantage of the high $p_{\perp}$ of their decay products. The decay channel $\rho\rightarrow Vh$ (with boosted Higgs) is a promising channel for the search for heavy vector resonances. In order to improve searches for resonances related to strong electroweak symmetry breaking a better understanding of possible interactions between vector and fermion resonances is needed. Such interactions can significantly modify couplings of $\rho$ resonances to SM quarks. Moreover, if $\rho$ decays into fermion resonances are kinematically allowed, they can easily dominate over all other decay channels and magnify significantly the overall decay width, making direct searches for spin-1 resonances at the LHC much more difficult.

\section*{Acknowledgments}

This work was supported by the Advanced Grant EFT4LHC of the European Research Council (ERC).

\end{document}